\documentclass{article}
\usepackage[utf8]{inputenc}
\usepackage{authblk}
\usepackage{booktabs}
\usepackage{graphicx}
\usepackage{multirow,amssymb}
\usepackage[table,xcdraw]{xcolor}
\usepackage{subcaption}
\usepackage{longtable}
\usepackage{lscape}

\newcommand\rotV[1]{\rlap{\rotatebox{90}{#1}}}

\title{A Survey on Business Process View Integration}

\author[1]{Rafael Belchior\thanks{rafael.belchior@tecnico.ulisboa.pt}}
\affil[1,2,3,4]{INESC-ID, Instituto Superior Técnico}

\author[2]{Sérgio Guerreiro}

\author[3]{André Vasconcelos}

\author[4]{Miguel Correia}

\begin{document}

\maketitle
\section*{Abstract}
The complexity of a business environment often causes organizations to produce several inconsistent views on the same business process, leading to fragmentation and inefficiencies.  
Business process view integration attempts to produce an integrated view from different views of the same model, facilitating the management of models. 

To study trends around business process view integration, we conduct a systematic literature review to summarize findings since the 1970s, up to its potential novel applications. With a starting corpus of 798 documents, this survey draws up a systematic inventory of solutions used in the academia and in the industry. By narrowing it down to 51 articles, we discuss in-depth 15 business process integration techniques papers. After that, we classify existing solutions according to their practicality. Our study shows that most integrated views are constructed by annotation, using formal merging rules.

Finally, we explore possible future research directions. We highlight the application of view integration to the blockchain research area, where stakeholders can have different views on the same blockchain. We expect that this study contributes to interdisciplinary research across view integration.


\section{Introduction}
\label{sec:intro}
A \emph{business process} (BP) is a collection of activities (or tasks), representing a well-defined procedure that aims to achieve a specific organizational goal \cite{saven2003}. Core assets of organizations, business processes shape the functioning and efficiency of organizations.

\emph{Business process models} represent business processes and aim at facilitating communication between stakeholders \cite{indulska2009}, serving as the initial point to guide business decisions. Business process models can be \emph{instantiated}, allowing for customization. Business process models are difficult to manage, sometimes accounting for several thousand models, especially if there are variations or different views of a given process. To manage such complexity, analysts can leverage \emph{business process management} (BPM) techniques \cite{Song2011,aalst2003}.

\begin{figure}[h]
    \centering
    \includegraphics[scale=0.4]{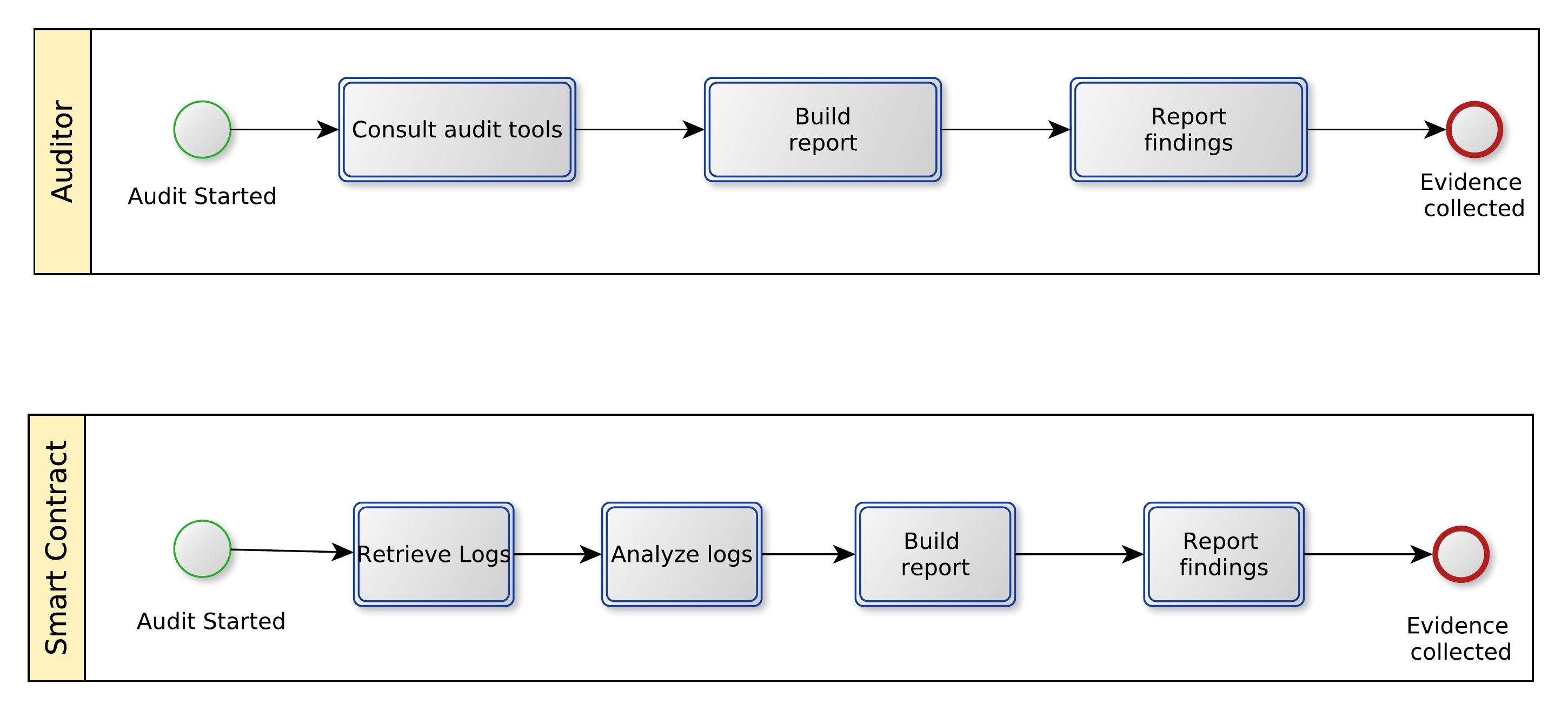}
    \caption{Auditor and Smart Contract (representing a blockchain consortium) concerns' regarding the collection of evidence for a semi-automatic audit}
    \label{fig:views}
\end{figure}

\emph{Business process management} (BPM) provides tools, methods to design, optimize, and maintain business processes.
BPM typically requires five steps: process identification, process discovery, process analysis, process redesign, and process implementation, monitoring, and controlling. These five-step-process outputs business processes that can be represented as different \emph{views} (e.g., organizational, stakeholder, information, application). Thus, one may have several views on the same business process. Figure \ref{fig:views} represents two different views of the business process: collecting evidence for semi-automated audits using blockchain \cite{belchior2019_audits,Belchior2019}.

Business processes are represented with \emph{business process modeling languages} (BPMLs), such as \emph{Event-driven process chains} (EPC), and \emph{Business Process Modeling Notation} (BPMN). EPCs are a representation of business processes flow charts
\cite{Gottschalk2008}. An EPC has three node types: events, functions, and logical connectors. Events are passive elements that constitute pre-requisites for the execution of functions. Logical connectors determine the process behavior, e.g., by associating two events or functions (via a directed arc). Connector types include XOR, $\land$ (and), and $\lor$ (or). 

Although EPCs are much used, the industry standard for representing business processes is the {Business Process Modeling Notation} (BPMN) \cite{Brocke2015,white2017}. BPMN aims to support business process management by providing a notation that can represent complex business semantics. BPMN defines flow nodes (events, activities, and gateways), connecting (sequences, messages, and associations), and swimlanes and artifacts.

\emph{Business process view integration} (BPVI), is the discipline studying the consolidation of different views regarding a {business process} \cite{Tran2011,Dongen2013,Dijkman2008b}. When we refer to view integration in this paper, by default we mean BPVI. However, this term can also refer to other integration techniques (e.g., database schema integration). Therefore, view integration is a superset of BPVI. The existence of different views stems from stakeholders conducting a business process in different ways. BPVI addresses three main challenges: 1) languages widely used for business process modeling are not adequate to promote the reuse of models \cite{Tran2011}, 2) while processes executed by one stakeholder are easy to document, processes involving stakeholders with different incentives and views are much more cumbersome \cite{colaco2017}, and 3) business modeling reflects the modeling team's perspective and a different team might come with a different representation \cite{colaco2017}. Therefore, human participation in business processes needs to be addressed \cite{holmes2008}.

BPVI processes create \emph{integrated views}. The concept of integrated view has its roots in database schema integration. Database schema integration is the set of activities that integrates the different schemas on a single, unified schema \cite{batini86}. In schema integration, a global conceptual description of a database is created. This concept influenced business analysts to perform the same on business processes.

This paper reviews the state-of-the-art regarding business process view integration and explores its trends. We, therefore, address the following \emph{research questions} (RQ): 
\begin{itemize}
    \item RQ1: 
    What is the origin of business process view integration, and what is its evolution?
    
    \item RQ2:
    What are the current business process view integration techniques in the literature, and what is their taxonomy?
    
    \item RQ3: What are the future trends of view integration?
\end{itemize}

To answer RQ1, we first elaborate on the past of business process view integration: database schema integration, providing an informal survey on papers dated from the 70s to around the 00s. After that, we refer to RQ2 by providing a systematic and comprehensive survey that reviews and classifies existing techniques for view integration. Finally, we argue that view integration can be applied to blockchain technology, answering RQ3.

This document is organized as follows: Section \ref{sec:vi} introduces view integration and presents this survey's methodology. After that, it elaborates on the classification criteria and the identified view integration techniques. Section \ref{sec:discussion} discusses the results. Section \ref{sec:future_trends} presents future trends for business process view integration, followed by the related work, in Section \ref{sec:rw}. Finally, Section \ref{sec:conc} concludes the paper.

\section{BPVI Systematic Literature Review}
\label{sec:vi}
In this section, we introduce the view integration research area. First, we provide historical framing by recalling database view integration. After that, we execute a systematic literature review on view integration regarding business process view integration.

\subsection{An Historical Perspective}
\label{subsec:past}

\begin{figure}[h]
    \centering
    \includegraphics[scale=0.26]{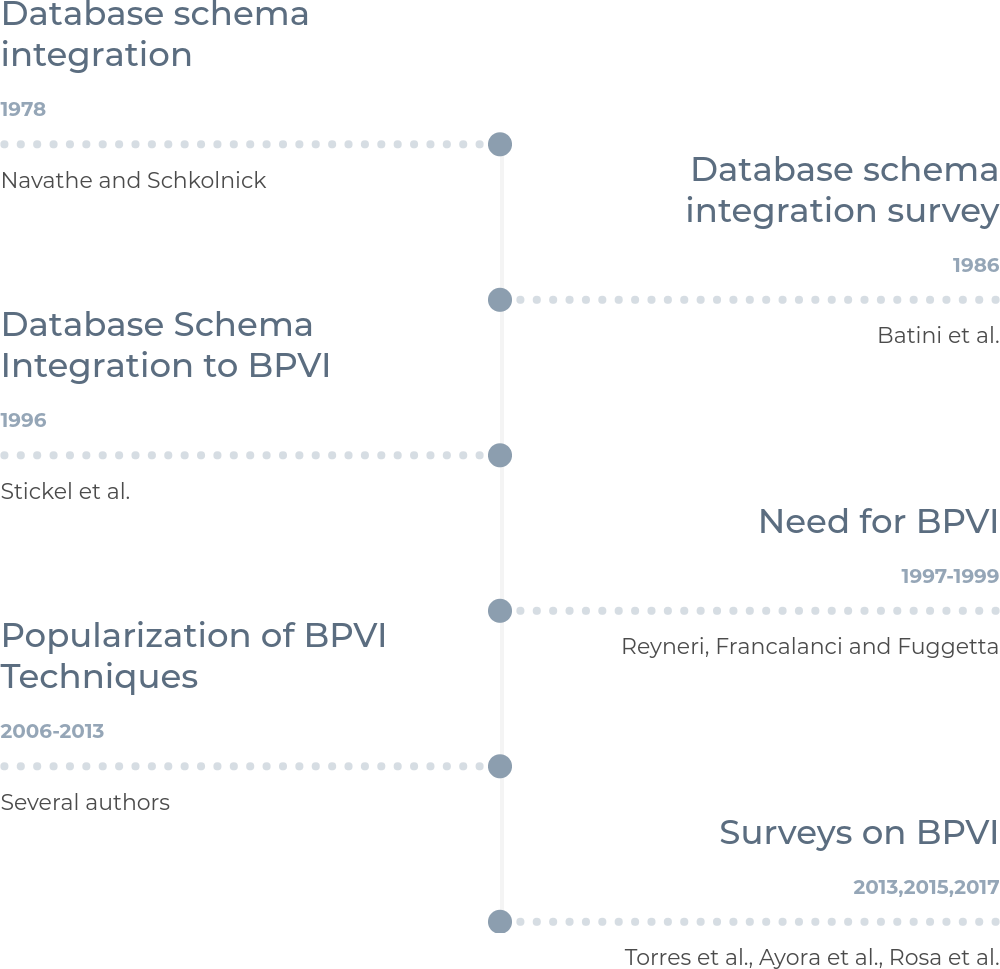}
    \caption{Historical perspective on BPVI}
    \label{fig:timeline2}
\end{figure}

\emph{Database view integration} is the research area that prompted the emergence of business process view integration. In the context of databases, the goal of view integration is to produce a holistic description of databases by combining the different database users' views. User views are collected and integrated, yielding the conceptual database schema. Past trends include view integration techniques applied to database schema integration, having its inception in the late 70s and popularized in the 80s. This historical perspective is illustrated in Figure \ref{fig:timeline2}. 

In database schema design, the general idea is to capture different stakeholders' views on the data that are identified and analyzed into several \emph{input schema}. After that, the input schemes are consolidated, based on their similarities, producing integrated schemas. Navathe and Schkolnick popularized the idea of database schema integration in 1978 \cite{navathe78}. The same authors present a conceptual framework for logical database design, including the view modeling and view integration steps, contributing to view integration in database design \cite{navathe78,navathe82}.

Later, in 1984, Dayal and Hwang expanded database schema integration to multiple, heterogeneous distributed databases \cite{Dayal84}. The authors describe a view definition schema applied to a functional data model resolving inconsistencies across heterogeneous databases, creating a consolidated view. View integration then became part of the larger database design activity, required to respond to the data requirements users have. These studies influenced future work on VI for object-oriented databases \cite{Gotthard92}.

In 1986, Batini et al.\ provided a systematic literature review on methodologies for database schema integration \cite{batini86}, comparing methodologies for database schema integration. In 1996, schema integration was considered a necessity to eliminate redundancies and maintain consistency across database systems \cite{stickel96}. The authors from the same study established the bridge between database schema integration and `` a business process-oriented strategy
for data integration''. Preuner and Schrefl study the integration of views of object life-cycles represented by behavior diagrams \cite{Preuner98}.

The last years of the 20th century then paved the way for the study of view integration techniques in business processes. Dijkman et al.\ highlighted ``that schema integration is not able to cope with heterogeneous control flow representation of BPM schemas''  \cite{Dijkman06}. Stumptner et al.\  \cite{Stumptner2004} defined consistency criteria for behavior integration. Vöhringer \cite{kcpm05} presented a parallel between schema integration and view integration, pointing out the some challenges, namely structure heterogeneity, and user communication. 

Many of these contributions fostered the transition of schema integration to view integration by applying concepts from databases to business processes. We now introduce the methodology for our survey on BPVI.

\subsection{Methodology}
\label{sec:methodology}

We applied the procedure proposed by Webster and Watson \cite{weber2002} and taking into account teachings from Briner and Denyer \cite{briner2012}. Thus, we divide our review methodology into the following steps: 

\begin{enumerate}
    \item  Identification of the research questions and the goals of the systematic literature review (RQ1, RQ2, RQ3).
    \item Preparation of a proposal and review protocols for the review.
    \item Search the literature for relevant studies addressing the research questions (study identification).
    \item Select the studies, critically appraise the study, take notes and summarize the collected information (study selection).
    \item  Disseminate the review findings.
\end{enumerate}

Step 1 corresponds to the definition of the research questions (Section \ref{sec:intro}). The present section implements the second step of the methodology.

\subsubsection*{Paper Inclusion Criteria}

The eligibility of the studies for this survey is based on predefined inclusion and exclusion criteria. Papers included in this survey satisfy at least one inclusion criteria: 1) the paper includes an exposition and discussion on BPVI techniques (in particular database schema integration or business process view integration), 2) the paper selected to be included in this review describes a technique that supports BPVI, 3) the paper provides relevant discussions to establish a bridge between BPVI and emerging research areas.

\subsubsection*{Study Identification}
To address the exposed research questions, we perform a systematic literature whose scope encompasses work up to July 2020. Firstly, we performed a keyword-based search on electronic databases. The following query strings to identify the relevant publications regarding BPVI: ``business process view integration", "business" AND "process" AND "view" AND "integration", "business process view" AND "integration", and "business process" AND "view integration". Due to the high number of search results the analysis of search results was skipped for most searches. Only the search with the keywords "business process" AND "view integration" was included, as it is the most specific. The first group of papers was obtained from the Google Scholar database, on April 2020, according to Table \ref{tab:slr_search_res}. One can see that Google Scholar indexes more literature than Science Direct, so we used the first library. We finished the search after 30 irrelevant publications. After that, a selection of the identified literature was conducted based on the abstract and titles. We searched for referenced works on those papers (called ``snowballing"), and also in different grey literature sources -- the process was repeated until theoretical saturation was reached (i.e., snowballing did not yield relevant results).

\begin{table}[h]
\centering
\caption{Literature review search results}
\label{tab:slr_search_res}
\resizebox{\textwidth}{!}{%
\begin{tabular}{@{}lll@{}}
\toprule
\textbf{Search Term |Scientific Database} & \textbf{Google Scholar} & \textbf{Science Direct} \\ \midrule
``business process view integration" & 5 & 0 \\
"business" OR "process" OR "view" OR "integration" & 4 070 000 & 163,382 \\
"business process view" AND "integration" & 904 & 54 \\
"business process" AND "view integration" & 788 & 62 \\ \bottomrule
\end{tabular}%
}
\end{table}


We remark that some of the work done in view integration is performed within the industry. To address that, and to reduce the publication bias, we included grey literature in our research, as advised by Mahood et al.\  \cite{greyliterature}. Publication bias stems from the fact that studies with statistically significant results (e.g., hypothesis corroborated by the authors) are more likely to be published and thus discovered in search processes. We thus define grey literature as academic theses, unpublished research, blog posts, and technical reports. To evaluate grey literature, additionally to the documents indexed by Google Scholar, we evaluated the first 100 hits from Google Search with the same keywords that we used on Google Scholar. Grey literature and other sources accounted for 10 studies.

\begin{figure}[]
    \centering
    \includegraphics[scale=0.3]{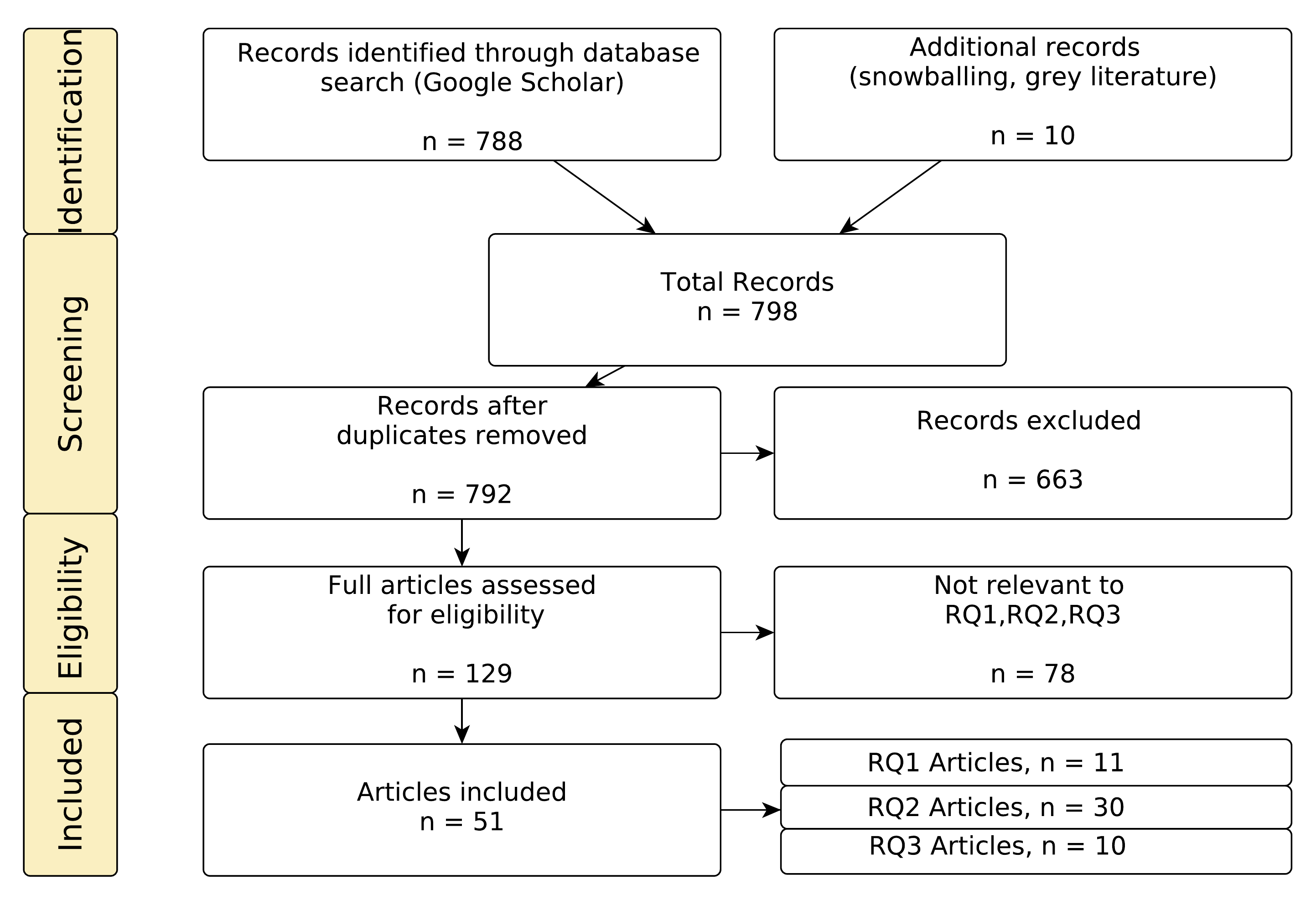}
    \caption{PRISMA diagram specifying our literature research methodology}
    \label{fig:prism}
\end{figure}

Initially, we retrieved 798 studies (788 from Google Scholar, and 10 from other sources). We removed the duplicates and obtained 792 studies. After that, we analyzed the title, abstract, and keywords, yielding 127 relevant studies. After the initial screening, full-text articles were assessed for eligibility; this yielded 15 studies directly related to BPVI, and 15 support studies (contributing to the area of BPVI), yielding a total of 30 articles that answer RQ2. A total of 21 articles were included to answer RQ2 and RQ3. Figure \ref{fig:prism} represents an adapted \emph{Preferred Reporting Items for Systematic Reviews and Meta-Analyses} (PRISMA) diagram \cite{prisma} considering all the steps of our literature research methodology. For a complete list of the papers responding to RQ1, RQ2, and RQ3, please refer to Appendix A.

\subsubsection*{Threats to Validity}
The main threats to the validity of the present survey are: \emph{language bias}, \emph{selection bias}, and \emph{the focus on view integration}.

Language bias refers to the fact that only studies in English have been included. Selection bias occurs because the search for papers focused on academic venues, including journal, conference, and workshop papers. Although we alleviate this threat by consulting grey literature, some relevant works may be left out.
Lastly, we state that the area of business process view integration is more encompassing than what we focus on in this survey. For instance, we deliberately leave out most work on consolidated models, a term related to integrated view. We illustrate how our study differentiates from others (and how the reader can obtain a completing perspective, accordingly) in Section \ref{sec:rw}.

\subsection{Classification criteria for view integration techniques}
\label{subsec:criteria}
In this section, we outline the categories and classification criteria for view integration techniques. We follow an adapted version of a taxonomy for business process variability modeling \cite{rosa2017}, explained in this section.

We classify each view integration solution in one of two categories. The first category, \emph{automation}, regards the business analyst's effort of setting up the view integration technique and has three values: manual, semi-automated, and automated. Manual automation happens when the analyst wants to merge two or more views, mostly manually (with a tool). Semi-automated methods rely on matching criteria, which allow decreasing some efforts from the analysts. The analyst matches parts of the views to be integrated, and the corresponding tool performs the integration. Automated methods do not rely on user input, merging all components of the views.

The second category, \emph{matching} technique, refers to how elements of a business process are linked to a predicate over properties of the application domain. In other words, it refers to how objects are linked with their correspondent objects on the other view. Matching is divided into \emph{annotations}, and \emph{behaviour}.

Annotations allow to systematically capture process knowledge, providing the necessary semantics for analysts to model processes. Using this concept, process performers can systematically capture process knowledge, and process engineers can incorporate it into process models for process model maintenance. Annotations are useful since they attribute semantics to certain objects of a business model, matching them with similar objects. Annotations are attached to model artifacts referring to implicitly defined elements of the discourse domain, typically using ontology-supported linguistic techniques, and can be matched with a similar one on a different view, creating a link. Behavior matching evaluates the effects or functionality of a business process, usually leveraging a domain ontology. In other words, two elements are linked if they produce the same results. 

Finally, we categorize solutions by their \emph{merging} technique. We label methods with \emph{formal} if the merging technique follows a well-defined algorithm, or \emph{adhoc} if the merging technique is not explicit or varies. This practical and summarized classification allows for assessing the practicality of each solution -- regarding its effort (automation), matching technique (annotation/behavior), and merging technique (formal/ad-hoc).

To further classify models, we explicit the following properties, based on a recent survey \cite{rosa2017}. In this survey, Rosa et al.\ propose a similar but different concept for view integration: consolidated model. A consolidated model is referring to the business process, whereby its elements originate from different processes. The consolidated model is semantically equivalent to the original processes and can be changed dynamically, conserving the relationships of its elements \cite{Morrison2009,rosa2010}. With this in mind, the consolidated model includes the concept of the integrated view.

For each technique we analyze the following aspects: 

\begin{enumerate}
    \item \emph{Language} criteria refers to the primary business process modeling language used. 
    \item \emph{Extension} encompasses a consolidated view that contains the behavior shared by all views. The consolidated view can be extended to produce a specific view.
    \item  \emph{Restriction} allows obtaining a variant of a consolidated view by enforcing restrictions on the original model (for example, skipping activities on the original model).
    \item \emph{Abstraction} criterion is fulfilled if a user can customize a view. For example, some approaches rely on ``annotations'', or another explicit linkage to provide semantics.
    \item  \emph{Structural} (correctness) assesses to if the tool can provide guarantees about the correctness of consolidated views (e.g., no isolated nodes).
    \item \emph{Behavioral} (correctness) assesses to if the tool can guarantee the correct behavior of the consolidated models (e.g., avoiding deadlocks).
    \item \emph{Formalization} defines if a method has concrete algorithms and/or definitions.
    \item  \emph{Implementation} criteria defines if a method is implemented.
    \item  \emph{Validation} criteria apply if a method was applied to a real-world scenario through discussions with domain experts. 
\end{enumerate}

\subsection{View Integration Overview}
\label{subsec: vit}
This section elaborates on the different view integration techniques, separated by the matching technique criteria: annotation or behavior.

\subsubsection{Matching by Annotation}
This section describes solutions whose matching method is ``annotation''.

In \cite{Mendling2006}, the authors present a view integration technique applied to Event-driven process chains.
The study introduces a merge operator that takes two EPCs and their semantic relationships as input and produces an integrated EPC. For that, semantic relationships have to be identified by a business process designer. Each pair of nodes describing the same real-world events is merged into a single node, and the former input and output arcs are joined and split with AND connectors, respectively. The arcs of each pair of nodes part of a sequence are refactored. Finally, a set of restructuring rules is proposed to eliminate unnecessary structure (i.e., reducing the resulting EPC size). After the annotation, the tool merges the models by following a specific set of rules.

Morrison et al.\ provide a theoretical framework for assessing the integration of business processes \cite{Morrison2009}, exemplifying its application to a family of processes merged via an ad-hoc method. The authors use clustering techniques to classify business processes. For example, k-mean clustering can be used to create clusters of business processes sharing common traits.  After that, the integration goals are defined (they can be provided by analysts or inferred based on each business process). The integration occurs where nodes and edges are added or removed from the model representing a pair of SPNets. The outcome of the integrations is assessed using similarity metrics.

Tran et al.\ \cite{Tran2011} propose a name-based matching approach for view integration, based on the \emph{view-based modelling framework} \cite{tran2007}. The proposed matching approach, name-based, is a  semi-automatic method that pairs the modeling entities by their name, which pose the same functionality and semantics. In this scheme, the business process analyst defines the business process using a custom framework, a view-based modeling framework. 
The main idea of name-based matching for view integration is to find all integration points (elements from different views that share the same name) between two views and merge those two views at the integration point. 

In \cite{colaco2017}, the authors propose an incremental approach to infer consolidated business process diagrams from different views, applied to BPMN 2.0. Their approach is based on previous work \cite{caetano2012, pereira2011,pereira2011b}, in which an organizational taxonomy is proposed by specifying six business process dimensions. A business process model repository has the ``time factor'' embedded, allowing time dependencies on the possibly various versions of a business model. The integration process begins with the modeling of a specific view of a process. The classification of each view is performed by the stakeholder while inputting information to the repository.

In \cite{huang2014}, BPMN process models are decomposed via decomposition techniques into fragments. The authors annotate each activity with its immediate effects and calculate the effect accumulation. 
Later, the E-RPST merging algorithm is applied, mapping nodes with its highest similarity score pair, yielding a consolidated model.

Derguech et al.\ \cite{derguech2017} propose an algorithm for merging process models into a configurable process model anchored on annotations, for capability-annotated process graphs, abstracting from BPMN, EPC, and other commonly used notations in this research area. The paper proposes an algorithm that inputs a set of capability-annotated process models and outputs a capability-annotated configurable model.

In \cite{kunchala2017}, the authors propose a method for merging collaborative inter-orga\-ni\-za\-tio\-nal business processes, providing a theoretical contribution to the types of merging techniques (synchronous vs.\ asynchronous, interactive vs.\ non-interactive). In subsequent work \cite{kunchala2019}, the authors generate artifact lifecycles from the activity-centric from the inter-organizational business processes. The proposed approach combines the nodes of collaborating processes to generate a consolidated process.

Rosa et al.'s work \cite{rosa2010,rosa2013} presents an algorithm that produces a unified, configurable business process model from two different ones.
This algorithm works with several representations of business models, such as EPC and BPMN, and leverages a merging operator. The first step of the merging algorithm is to process business process models into configurable process graphs.

In \cite{cardoso2020}, Cardoso and Sousa propose an approach that follows previous work \cite{colaco2017, sousa2019}, and generates stakeholder-specific models using functional decomposition by recursively breaking down a process as sub-activities. Such models are generated from a consolidated model and user input in the six Zachaman dimensions: what, when, how, why, who, and where. The ``What'' refers to the enterprise's information, focusing on data. ``When'' expresses how an artifact evolves with the timeline, focusing on time. ``How'' tackles the execution of the enterprise's mission, focusing on the function. ``Why'' translates motivation into objectives. ``Who'' indicates people behind business operations. Finally, ``Where'' refers to the enterprises' artifacts' distribution, focusing on the network

\subsubsection{Matching by Behavior}
This section describes solutions whose matching method is ``behavior''.

Grossman et al.\ propose integration operators to create, manage, and finalize composition between autonomous object-oriented systems \cite{Grossmann2005}. A structured sequence of integration steps that analyzes the relationships between processes to be integrated are presented. After that, a set of integration options is proposed that can specify a high-level integration operator that conducts the integration.

Gottschalk et al.\ present a three-phase approach that merges two business process models, represented by an Event-driven Process Chains \cite{Gottschalk2008} into a consolidated model. First, the EPCs to be merged are reduced to function-graphs, entities that depict the active behavior of the EPCs. Secondly, the two function-graphs are reduced to a single, consolidated one. Finally, the consolidated function-graph is transformed into an EPC. 

Kuster et al.\ \cite{kuster2008} present a simple prototype focused on business process development. In this paper, differences between process models to be merged are detected using correspondences between model elements and the technique of Single-Entry-Single-Exit fragments (SESE fragments). For each detected difference, a resolution transformation is generated, merging the models.

Li et al.\ \cite{Li2009} propose an algorithm to output a consolidated model by conducting a heuristic search across the process graph (represented in ADEPT), using a measuring distance to find reference models with minimal average weighted distance to the variants.

In \cite{Assy13}, the authors merge process fragments around BPMN activities to construct a consolidated fragment for each activity, instead of merging whole process models. The authors abstract some BPMN concepts to build a notation graph.

\section{Discussion}
\label{sec:discussion}

This section discusses and classifies view integration techniques, as defined in Section \ref{subsec:criteria}.

\subsection{Overview of View Integration Approaches}

Issues on the integration of business processes \cite{francalanci1997} and reusability of business process models started to be tackled as early as 1997 and 1999 \cite{carla1999}, respectively. Early research suggested that a business process would not need to be redesigned from scratch every time a model is modified by employing reusable building blocks.

Table \ref{tab:bpvi} depicts the comparison between the various view integration techniques.  We present the main techniques for database view integration, in Section \ref{subsec:past}, as a contextualization of BPVI techniques. From the present table, one can see that most of the available solutions are semi-automatic solutions relying on annotations to produce an integrated view.

\begin{table}[]
\centering
\caption{Evaluation of business process view integration methods, ordered by year of publication.}
\label{tab:bpvi}
\resizebox{\textwidth}{!}{%
\begin{tabular}{@{}lcccccccccccccc@{}}
 &
  \multicolumn{1}{l}{} &
  \multicolumn{1}{l}{} &
  \multicolumn{1}{l}{} &
  \multicolumn{1}{l}{} &
  \multicolumn{1}{l}{} &
  \multicolumn{1}{l}{} &
  \multicolumn{2}{c}{} &
  \multicolumn{3}{c}{} &
  \multicolumn{3}{c}{} \\
 &
  \multicolumn{1}{l}{} &
  \multicolumn{1}{l}{} &
  \multicolumn{1}{l}{} &
  \multicolumn{1}{l}{} &
  \multicolumn{1}{l}{} &
  \multicolumn{1}{l}{} &
  \multicolumn{2}{c}{\multirow{-2}{*}{}} &
  \multicolumn{3}{c}{\multirow{-2}{*}{}} &
  \multicolumn{3}{c}{\multirow{-2}{*}{}} \\
\multicolumn{1}{c}{\rotV{Reference}} &
  \rotV{Year} &
  \rotV{Citations} &
  \rotV{Automation} &
  \multicolumn{1}{l}{\rotV{Matching}} &
  \multicolumn{1}{l}{\rotV{Merging}} &
  \rotV{Language} &
  \rotV{Extension} &
  \rotV{Restriction} &
  \rotV{Abstraction} &
  \rotV{Structural} &
  \rotV{Behavioral} &
  \rotV{Formalization} &
  \rotV{Implementation} &
  \rotV{Validation} \\ \midrule
\cite{Grossmann2005} &
  2005 &
  39 &
  SA &
  Behavior &
  FMR &
  UML &
  \cellcolor[HTML]{D8E3BB}+ &
  \cellcolor[HTML]{BF504D}- &
  \cellcolor[HTML]{BF504D}- &
  \cellcolor[HTML]{D8E3BB}+ &
  \cellcolor[HTML]{D8E3BB}+ &
  \cellcolor[HTML]{D8E3BB}+ &
  \cellcolor[HTML]{BF504D}- &
  \cellcolor[HTML]{BF504D}- \\ \midrule
\cite{Mendling2006} &
  2006 &
  121 &
  SA &
  Annotation &
  FMR &
  EPC &
  \cellcolor[HTML]{D8E3BB}+ &
  \cellcolor[HTML]{BF504D}- &
  \cellcolor[HTML]{BF504D}- &
  \cellcolor[HTML]{D8E3BB}+ &
  \cellcolor[HTML]{D8E3BB}+ &
  \cellcolor[HTML]{D8E3BB}+ &
  \cellcolor[HTML]{BF504D}- &
  \cellcolor[HTML]{BF504D}- \\ \midrule
\cite{tran2007} &
  2007 &
  56 &
  SA &
  Annotation &
  FMR &
  BPEL &
  \cellcolor[HTML]{D8E3BB}+ &
  \cellcolor[HTML]{BF504D}- &
  \cellcolor[HTML]{BF504D}- &
  \cellcolor[HTML]{D8E3BB}+ &
  \cellcolor[HTML]{D8E3BB}+ &
  \cellcolor[HTML]{D8E3BB}+ &
  \cellcolor[HTML]{D8E3BB}+ &
  \cellcolor[HTML]{BF504D}- \\ \midrule
\cite{Gottschalk2008} &
  2008 &
  107 &
  A &
  Behavior &
  FMR &
  EPC &
  \cellcolor[HTML]{D8E3BB}+ &
  \cellcolor[HTML]{BF504D}- &
  \cellcolor[HTML]{BF504D}- &
  \cellcolor[HTML]{D8E3BB}+ &
  \cellcolor[HTML]{D8E3BB}+ &
  \cellcolor[HTML]{D8E3BB}+ &
  \cellcolor[HTML]{D8E3BB}+ &
  \cellcolor[HTML]{BF504D}- \\ \cmidrule(r){1-14}
\cite{Li2009} &
  2009 &
  115 &
  A &
  Behavior &
  FMR &
  ADEPT &
  \cellcolor[HTML]{D8E3BB}+ &
  \cellcolor[HTML]{D8E3BB}+ &
  \cellcolor[HTML]{D8E3BB}+ &
  \cellcolor[HTML]{D8E3BB}+ &
  \cellcolor[HTML]{BF504D}- &
  \cellcolor[HTML]{D8E3BB}+ &
  \cellcolor[HTML]{D8E3BB}+ &
  \cellcolor[HTML]{BF504D}- \\ \midrule
\cite{Morrison2009} &
  2009 &
  28 &
  SA &
  Annotation &
  Ad-hoc &
  EPC (1) &
  \cellcolor[HTML]{D8E3BB}+ &
  \cellcolor[HTML]{BF504D}- &
  \cellcolor[HTML]{D8E3BB}+ &
  \cellcolor[HTML]{D8E3BB}+ &
  \cellcolor[HTML]{BF504D}- &
  \cellcolor[HTML]{D8E3BB}+ &
  \cellcolor[HTML]{BF504D}- &
  \cellcolor[HTML]{BF504D}- \\ \midrule
\cite{rosa2010} &
  2010 &
  137 &
  SA &
  Annotation &
  FMR &
  EPC &
  \cellcolor[HTML]{D8E3BB}+ &
  \cellcolor[HTML]{BF504D}- &
  \cellcolor[HTML]{D8E3BB}+ &
  \cellcolor[HTML]{D8E3BB}+ &
  \cellcolor[HTML]{D8E3BB}+ &
  \cellcolor[HTML]{D8E3BB}+ &
  \cellcolor[HTML]{D8E3BB}+ &
  \cellcolor[HTML]{BF504D}- \\ \midrule
\cite{Tran2011} &
  2011 &
  11 &
  SA &
  Annotation &
  FMR &
  VbMF (2) &
  \cellcolor[HTML]{D8E3BB}+ &
  \cellcolor[HTML]{BF504D}- &
  \cellcolor[HTML]{D8E3BB}+ &
  \cellcolor[HTML]{D8E3BB}+ &
  \cellcolor[HTML]{D8E3BB}+ &
  \cellcolor[HTML]{D8E3BB}+ &
  \cellcolor[HTML]{D8E3BB}+ &
  \cellcolor[HTML]{D8E3BB}+ \\ \midrule
\cite{rosa2013} &
  2013 &
  241 &
  SA &
  Annotation &
  FMR &
  EPC &
  \cellcolor[HTML]{D8E3BB}+ &
  \cellcolor[HTML]{BF504D}- &
  \cellcolor[HTML]{D8E3BB}+ &
  \cellcolor[HTML]{D8E3BB}+ &
  \cellcolor[HTML]{D8E3BB}+ &
  \cellcolor[HTML]{D8E3BB}+ &
  \cellcolor[HTML]{D8E3BB}+ &
  \cellcolor[HTML]{D8E3BB}+ \\ \midrule
\cite{Assy13} &
  2013 &
  12 &
  SA &
  Behavior &
  FMR &
  BPMN &
  \cellcolor[HTML]{D8E3BB}+ &
  \cellcolor[HTML]{BF504D}- &
  \cellcolor[HTML]{BF504D}- &
  \cellcolor[HTML]{D8E3BB}+ &
  \cellcolor[HTML]{D8E3BB}+ &
  \cellcolor[HTML]{D8E3BB}+ &
  \cellcolor[HTML]{D8E3BB}+ &
  \cellcolor[HTML]{D8E3BB}+ \\ \midrule
\cite{huang2014} &
  2014 &
  5 &
  SA &
  Annotation &
  FMR &
  BPMN &
  \cellcolor[HTML]{D8E3BB}+ &
  \cellcolor[HTML]{BF504D}- &
  \cellcolor[HTML]{D8E3BB}+ &
  \cellcolor[HTML]{D8E3BB}+ &
  \cellcolor[HTML]{D8E3BB}+ &
  \cellcolor[HTML]{D8E3BB}+ &
  \cellcolor[HTML]{D8E3BB}+ &
  \cellcolor[HTML]{BF504D}- \\ \midrule
\cite{colaco2017} &
  2017 &
  3 &
  SA &
  Annotation &
  Ad-hoc &
  BPMN &
  \cellcolor[HTML]{D8E3BB}+ &
  \cellcolor[HTML]{BF504D}- &
  \cellcolor[HTML]{D8E3BB}+ &
  \cellcolor[HTML]{BF504D}- &
  \cellcolor[HTML]{BF504D}- &
  \cellcolor[HTML]{BF504D}- &
  \cellcolor[HTML]{BF504D}- &
  \cellcolor[HTML]{D8E3BB}+ \\ \midrule
\cite{kunchala2017} &
  2017 &
  2 &
  SA &
  Annotation &
  FMR &
  BPMN &
  \cellcolor[HTML]{D8E3BB}+ &
  \cellcolor[HTML]{D8E3BB}+ &
  \cellcolor[HTML]{D8E3BB}+ &
  \cellcolor[HTML]{D8E3BB}+ &
  \cellcolor[HTML]{BF504D}- &
  \cellcolor[HTML]{BF504D}- &
  \cellcolor[HTML]{BF504D}- &
  \cellcolor[HTML]{BF504D}- \\ \midrule
\cite{derguech2017} &
  2017 &
  6 &
  SA &
  Annotation &
  FMR &
  Capability-Annotated &
  \cellcolor[HTML]{D8E3BB}+ &
  \cellcolor[HTML]{BF504D}- &
  \cellcolor[HTML]{D8E3BB}+ &
  \cellcolor[HTML]{D8E3BB}+ &
  \cellcolor[HTML]{D8E3BB}+ &
  \cellcolor[HTML]{D8E3BB}+ &
  \cellcolor[HTML]{D8E3BB}+ &
  \cellcolor[HTML]{D8E3BB}+ \\ \midrule
\cite{cardoso2020} &
  2020 &
  2 &
  SA &
  Annotation &
  Ad-hoc &
  BPMN &
  \cellcolor[HTML]{D8E3BB}+ &
  \cellcolor[HTML]{D8E3BB}+ &
  \cellcolor[HTML]{D8E3BB}+ &
  \cellcolor[HTML]{BF504D}- &
  \cellcolor[HTML]{BF504D}- &
  \cellcolor[HTML]{BF504D}- &
  \cellcolor[HTML]{D8E3BB}+ &
  \cellcolor[HTML]{D8E3BB}+ \\ \midrule
 &
  \multicolumn{1}{l}{} &
  \multicolumn{1}{l}{} &
  \multicolumn{1}{l}{} &
  \multicolumn{1}{l}{} &
  \multicolumn{1}{l}{} &
  \multicolumn{1}{l}{} &
  \multicolumn{1}{l}{} &
  \multicolumn{1}{l}{} &
  \multicolumn{1}{l}{} &
  \multicolumn{1}{l}{} &
  \multicolumn{1}{l}{} &
  \multicolumn{1}{l}{} &
  \multicolumn{1}{l}{} &
  \multicolumn{1}{l}{} \\
\multicolumn{5}{l}{Legend} &
  \multicolumn{1}{l}{} &
  \multicolumn{1}{l}{} &
  \multicolumn{1}{l}{} &
  \multicolumn{1}{l}{} &
  \multicolumn{1}{l}{} &
  \multicolumn{1}{l}{} &
  \multicolumn{1}{l}{} &
  \multicolumn{1}{l}{} &
  \multicolumn{1}{l}{} &
  \multicolumn{1}{l}{} \\
\multicolumn{13}{l}{\begin{tabular}[c]{@{}l@{}}A - Automated; \\ SA - Semi Automated; \\ FMR - Formal Merging Rules\end{tabular}} &
  \multicolumn{1}{l}{} &
  \multicolumn{1}{l}{} \\
\multicolumn{7}{l}{1 - Supports arbitrary languages} &
  \multicolumn{1}{l}{} &
  \multicolumn{1}{l}{} &
  \multicolumn{1}{l}{} &
  \multicolumn{1}{l}{} &
  \multicolumn{1}{l}{} &
  \multicolumn{1}{l}{} &
  \multicolumn{1}{l}{} &
  \multicolumn{1}{l}{} \\
\multicolumn{13}{l}{2 - An abstraction supporting several models such as BPMN, EPC, and UML Activity} &
  \multicolumn{1}{l}{} &
  \multicolumn{1}{l}{} \\
\multicolumn{1}{c}{\cellcolor[HTML]{D8E3BB}+} &
  \multicolumn{7}{l}{Criterion fulfilled} &
  \multicolumn{1}{l}{} &
  \multicolumn{1}{l}{} &
  \multicolumn{1}{l}{} &
  \multicolumn{1}{l}{} &
  \multicolumn{1}{l}{} &
  \multicolumn{1}{l}{} &
  \multicolumn{1}{l}{} \\
\multicolumn{1}{c}{\cellcolor[HTML]{BF504D}-} &
  \multicolumn{8}{l}{Criterion not fullilled} &
  \multicolumn{1}{l}{} &
  \multicolumn{1}{l}{} &
  \multicolumn{1}{l}{} &
  \multicolumn{1}{l}{} &
  \multicolumn{1}{l}{} &
  \multicolumn{1}{l}{}
\end{tabular}%
}
\end{table}

Formal merging rules are used as a matching mechanism, along with annotations. Behavior-based matching solutions, such as \cite{Gottschalk2008, Morrison2009} utilize clustering mechanisms, thus allowing for a more automated approach. While a greedy generation of integrated views can reduce manual labour, typically it is not as accurate as annotation-based methods. Typical languages that are used to integrate views are EPC and BPMN. The majority of the solutions allow constructing an integrated view by extension (vs. a minority allowing to build by restriction). This is because often the starting point for studying business process variability are standalone views that are merged and combined into an integrated view, and not a consolidated view that derives specialized views. Structural and behavioral guarantees can be given by the formal generation of integrated views, semantically equivalent to the original ones (distinguishable based on a given predicate). While most solutions have a formalization available, only approximately half are validated with a real-world use case scenario. Implementations, even if just proof of concepts, have been provided.

Next, we elaborate on work that directly or indirectly supports the discussed BPVI techniques.

\subsection{Supporting Studies}
\label{sec:support}
Some studies indirectly support business process view integration, such as managing view workflows \cite{Nguyen17,colaco2017}. Schumm et al.\ \cite{Schumm2010} present a meta-model for process business process views and illustrate the elementary process viewing patterns. The authors provide an implementation that supports the alteration pattern for modification of attributes and their values, applied to the BPEL specification. Another example is Weidlich et al.\ \cite{Weidlich2011}, who define a set-algebra for behavioral profiles to identify redundancies across business process models. Kuster et al.\ \cite{kuster2007} present the notion of compliance of a business process model with an object life cycle. The generality of the solutions is typically reduced: most solutions are tied to a specific implementation, although some authors provide a technology-agnostic model \cite{Mendling2006, rosa2010}.

The study of consistency across business process models was also relevant to the advance of the area \cite{Weidlich12}. Weidlich and Mendling studied control flow aspects and consistency notions. The authors concluded that the perception of consistency is tied to the behavioral equivalence (if two processes have the same behavior, they are consistent). Other authors studied horizontal business process model integration by formalizing semantics using abstract state machines \cite{schewe2015}.

Later, several authors \cite{milani2015} studied how different decomposition heuristics affect process model understandability and maintainability, concluding that ``no comparable consent regarding the question of how to decompose a process model''. Along with process visualization techniques for multi-perspective process comparisons \cite{pini2015}, the way has been paved for the settlement of the view integration area.

There are some efforts to classify the quality of the integration of business processes.  Morrison et al.\ provide one of the first theoretical frameworks for assessing the integration quality of business processes \cite{Morrison2009}. Morrison et al.\ establish a distance measure between two SPNets to ensure the consolidated model does not deviate considerably from the originating models. Others propose an algorithm to output a consolidated model by learning from a collection of (block-structured) process variants, using heuristics \cite{Li2009}, and evaluate such models. Process model repositories  \cite{apromore2009,colaco2017} allow business analysts to manage a large number of models for analyzing, visualizing, transforming, and creating customizable process models. In particular, Atlas allows to consolidate and derive views from a consolidated model.  

In short, nowadays, business process modeling has undergone a substantial improvement over the last twenty years, being aligned with good practices on sharing, reusing, optimizing, and specializing business process models, leading to better performance within organizations where several views on business processes exist. However, there are gaps in this research area: 1) empirical comparisons between view integration models and tools, 2) lack of models and tools to support the full lifecycle of integrated views, 3) and the application of this research area to several domains. In light of the presented challenges, we elaborate on how BPVI can be extended to other research areas.

\section{Future Trends}
\label{sec:future_trends}

We discuss the possibility of leveraging view integration research in the blockchain area. Blockchain is an emerging technology that secures decentralized, immutable, append-only data storage. On top of such secure storage, a computing framework can be maintained by a network of untrusted participants (or nodes) via smart contracts \cite{Belchior2020}. Nodes hold a replica of this data structure locally (called the ledger), agreeing on the next global state via a consensus mechanism. Changes to the global state are done via transactions, calls issued against a program running on the blockchain (called a smart contract). Thus, blockchain is used where stakeholders/organizations do not trust each other, suitable for cross-organizational interactions.  

Blockchain can be a supporting infrastructure of business processes, not only accounting for its decentralized execution, but also for a multitude of other use cases \cite{Belchior2020}. With this into mind, one can analyze a blockchain with the six Zachaman's dimensions \cite{zachaman2010,sousa2007}, as Figure \ref{fig:bc_state} shows. The Zachman Framework is a framework for describing an enterprise architecture, using six dimensions.

By using smart contracts as the core functionality provider of blockchains, Zachman's framework provides us hints to understand how view integration can be applied to blockchain. Each dimension is as follows:

\begin{figure}[h]
    \centering
    \includegraphics[scale=0.3]{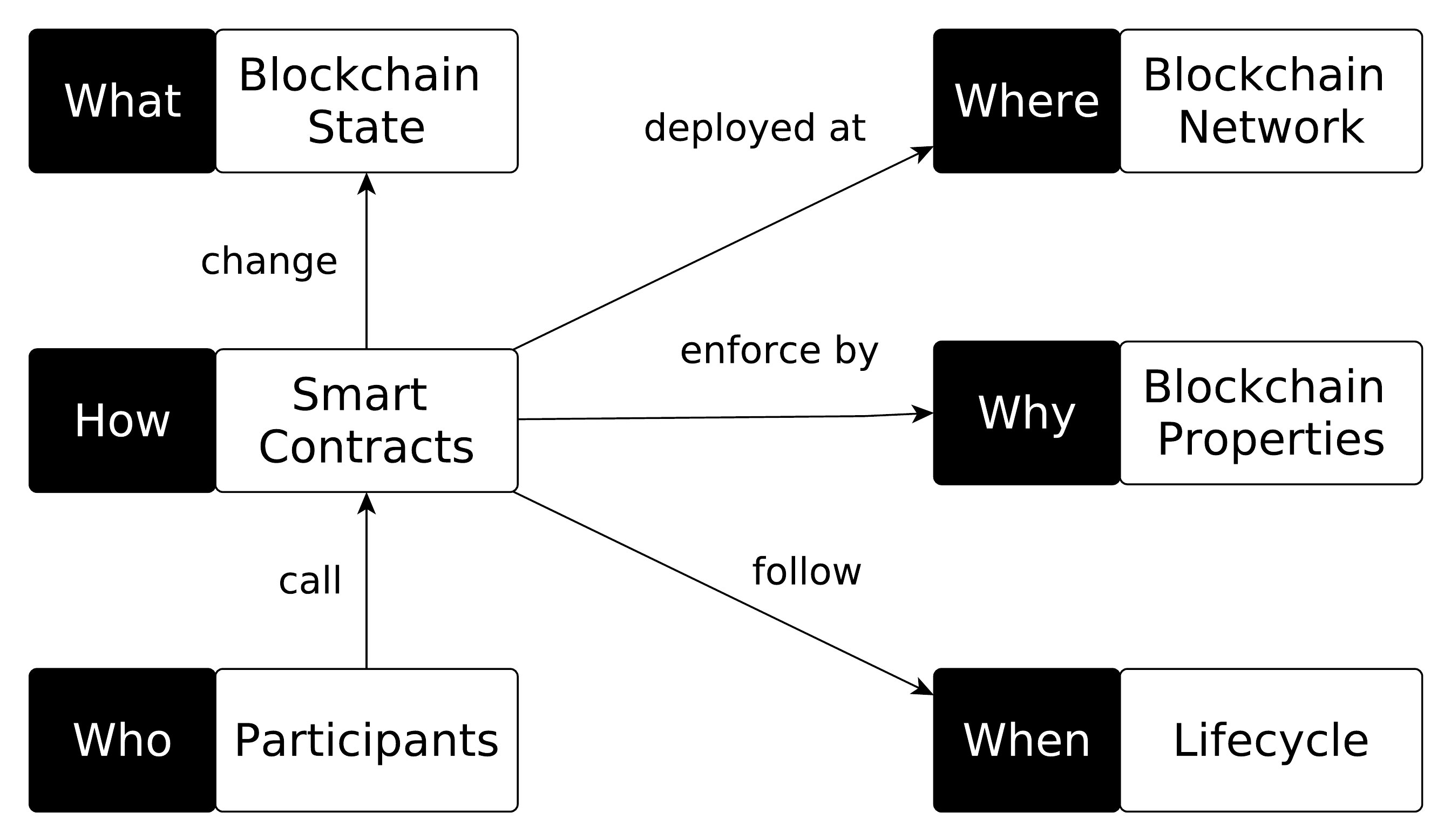}
    \caption{Representation of a blockchain state}
    \label{fig:bc_state}
\end{figure}

\begin{enumerate}
   \item What: corresponds to the data managed by an enterprise. The data corresponds to the blockchain state that changes according.
   
      \item Where: in which blockchain and in which node (and with which configuration) does the transaction take place.
      
    \item Who: the entities that have active behavior. Correspond to stakeholders submitting a transaction to a smart contract.
    
  \item How: translates organizational goals into its business. Smart contracts enforce a functionality and thus do that translation.
 
    \item When: expresses how each artifact evolves with the timeline. The timeline corresponds to the lifecycle of a smart contract. 
    
   \item Why: corresponds to the desired goals of a business process. A blockchain provides immutability, transparency, and traceability.
   
\end{enumerate}

Although a particular smart contract follows a specific lifecycle (When), on a specific blockchain (Where), following specific rules (How), the state accessed can differ (What), depending on the stakeholder accessing it (Who). A blockchain -- namely its data (states) and functionality (smart contracts)- can represent enterprises' concerns. Enterprise concerns vary according to its stakeholders - thus, local views and a global view exist.

Different views on the global state of blockchains can exist due to the blockchain's nature or data privacy necessity. For instance, in Bitcoin \cite{bitcoin}, the first public blockchain, the consensus is probabilistic, meaning that temporary views can exist.  Concurrently, on private blockchains tailored for enterprises, different views are not only common but desirable. For example, at Hyperledger Fabric, a popular enterprise-grade blockchain, the private data feature allows participants to hide part of the state they hold effectively, only sharing proof of the existence of such data \cite{fabric,hyperldegerpd}. Enterprise needs lead to the (permanent) existence of different views in the same blockchain.

A practical case for studying view integration on the blockchain is merging different blockchain views onto an integrated one \cite{Belchior2020}. This process has several real-world applications: i) \emph{blockchain migration}, ii) \emph{blockchain backup}, and iii) \emph{data analytics for legal audits}. In particular, blockchain migration happens when it is desirable to change a decentralized application's blockchain infrastructure to another. Setting up the new infrastructure requires an integrated view of the first blockchain. Regarding backups, while it may be desirable to save all the data containing the blockchain, perhaps a subset of that data (view) is enough. Concerning data analytics,  legal frameworks are starting to regulate blockchains \cite{eu19}, putting the focus on audits\footnote{https://www.coindesk.com/quadriga-kroll-analytics}\cite{audits2020, belchior2019_audits,Belchior2019}. Subsequently, it is important not only to have raw data but also to access each stakeholder's views to conduct a virtuous audit and analysis. Additionally, as views can contain sensitive information, view management would benefit from access control, e.g., blockchain-based \cite{Rouhani2020,ssibac}. Access control assures that only the entitled stakeholders can access and manipulate their views.

\begin{figure}[h]
    \centering
    \includegraphics[scale=0.4]{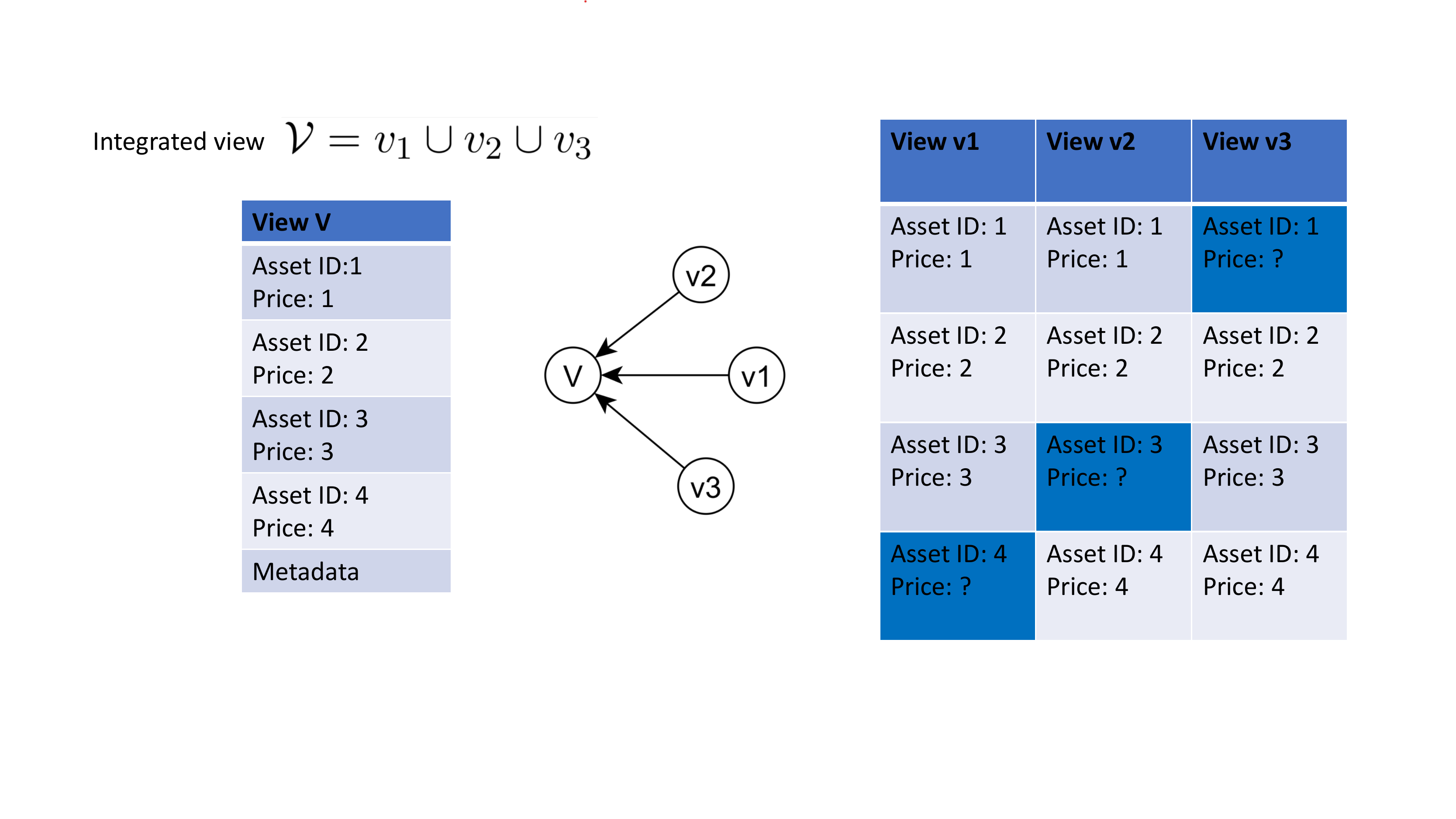}
    \caption{Blockchain View integration}
    \label{fig:bc_vi}
\end{figure}

In short, blockchain is often applied to mediate conflicting stakeholder concerns, allowing them to create their views, but at the same time, it provides a generic, integrated view \cite{Belchior2020}. Typically, as no entity fully controls the blockchain, the process of creating a consolidated view is manual and cumbersome. Future research is needed to systematically analyze those views, as well as automatically retrieve the consolidated view.

\section{Related Work}
\label{sec:rw}

Valença et al.\ analyze the literature around business process variability approaches and build a theoretical foundation around them, indicating the main challenges of the same research area  \cite{valenca2013}. Similarly, Mechrez and Reinhartz-Berger \cite{mechrez2014} investigate business process variability challenges but are more detailed in comparing and describing the solutions.

Torres et al.\ \cite{torres2013}, and Dohring et al.\ \cite{Dohring2014} focus on some business process variability approaches and comparing them by leveraging cognitive psychology concepts and empirical evaluation methods, respectively. Torres et al.\ compare behavioral approaches against structural approaches, providing a primer regarding the understability of such approaches. Dohring et al.\ conclude that ``we could not show that complexity or the participant's professional level significantly impacts the task success rate or user contentment''.

On \cite{ayora2015}, the authors propose the VIVACE framework, an empirical framework that allows assessing the expressiveness of a process modeling language about process variability approaches. The framework focuses on enabling process variability along the entire process lifecycle, contrasting with our study, which focuses on view integration.

Rosa et al.\ \cite{rosa2017} cover the most relevant literature on business process variability modeling. In this survey, the authors study process variability modeling. Several categories are introduced: node configuration, element annotation, activity specialization, and fragment customization approaches, based on the following criteria: process perspective, process type, customization type, supporting techniques, and extra-functional.  We partially rely on these classification criteria for classifying view integration techniques, as it is validated and well-validated by the scientific community. However, we conduct further updates from 2017 until September 2020 and present techniques directly applicable to view integration, a more focused scope than business modeling variability processing.

In short, our survey builds on top of previous work: our classification is practical rather than theoretical by representing the overall process conducting to view integration. We elaborate on the history of business process view integration, providing contextual insights that can guide future developments on the area. Hence, our work serves to pave the way for future trends, where we highlight the application of view integration to blockchain research.

\section{Conclusion}
\label{sec:conc}

In this survey, we have put into evidence the heterogeneity of solutions addressing (business process) view integration. For that, we conducted a systematic literature review, focusing not only on business process view integration but also its predecessor, database view integration. 

We introduce the historical area of database view integration that supports business process integration by analyzing its most influential literature. After that, we elaborate on three research questions from the analysis of 51 papers out of the original 798 documents: what is the origin of business process view integration, what are its current solutions, and what are its future trends. We found out that this area is maturing quickly, given the existence of robust tools, and that most view integration solutions use annotations and formal matching rules to create integrated views, as shown by Table \ref{tab:bpvi}. Moreover, some solutions offer behavioral and structural guarantees, providing a formalization of the view integration algorithms. However, empirical comparisions between view integration models and tools are still missing; and applications to other research areas can be done. 

Our survey paves the way for opening the discussion of applying view integration to decentralized ledger technologies, contributing to the advance of legal frameworks, operations, and management. In particular, the concept of blockchain view, and its management have practical applications for both the academia and the industry.

\section*{Acknowledgement}
 This work was supported by the EC through project 822404 (QualiChain), an
by national funds through Fundação para a Ciência e a Tecnologia (FCT) with reference UIDB/50021/2020 (INESC-ID).

\bibliographystyle{abbrv}
\bibliography{references2}

\appendix
\addcontentsline{toc}{section}{Appendix~\ref{app:scripts}: Training Scripts}
\section*{Appendix A - Included Studies}
\label{app:scripts}
This appendix contains the included articles on this survey, as well as their author, publication forum, and year of publication. 

\begin{table}[h]
\huge
\caption{Studies answering RQ1, RQ2, and RQ3}
\label{tab:my-table}
\resizebox{\textwidth}{!}{%
\begin{tabular}{@{}llllll@{}}
\toprule
\# &
  Reference &
  Year &
  Authors &
  Title &
  Publication Forum \\ \midrule
\multicolumn{1}{|l|}{1} &
  \multicolumn{1}{l|}{\cite{navathe78}} &
  \multicolumn{1}{l|}{1978} &
  \multicolumn{1}{l|}{Navathe, S and Schkolnick, M} &
  \multicolumn{1}{l|}{View representation in logical database design} &
  \multicolumn{1}{l|}{ACM SIGMOD international conference on management of data} \\ \midrule
\multicolumn{1}{|l|}{2} &
  \multicolumn{1}{l|}{\cite{Dayal84}} &
  \multicolumn{1}{l|}{1984} &
  \multicolumn{1}{l|}{Dayal, U and Hwang, H} &
  \multicolumn{1}{l|}{View Definition and Generalization for Database Integration in a Multidatabase System} &
  \multicolumn{1}{l|}{IEEE Transactions on Software Engineering} \\ \midrule
\multicolumn{1}{|l|}{3} &
  \multicolumn{1}{l|}{\cite{Gotthard92}} &
  \multicolumn{1}{l|}{1992} &
  \multicolumn{1}{l|}{\begin{tabular}[c]{@{}l@{}}Gotthard, W and Lockemann, P\\  and Neufeld, A\end{tabular}} &
  \multicolumn{1}{l|}{System-Guided View Integration for Object-Oriented Databases} &
  \multicolumn{1}{l|}{IEEE Transactions on Knowledge and Data Engineering} \\ \midrule
\multicolumn{1}{|l|}{4} &
  \multicolumn{1}{l|}{\cite{batini86}} &
  \multicolumn{1}{l|}{1986} &
  \multicolumn{1}{l|}{Batini, C. and Lenzerini, M. and Navathe, S.} &
  \multicolumn{1}{l|}{A comparative analysis of methodologies for database schema integration} &
  \multicolumn{1}{l|}{ACM Computing Surveys} \\ \midrule
\multicolumn{1}{|l|}{5} &
  \multicolumn{1}{l|}{\cite{stickel96}} &
  \multicolumn{1}{l|}{1996} &
  \multicolumn{1}{l|}{Stickel, E and Hunstock, J and Ortmann, A} &
  \multicolumn{1}{l|}{A Business Process Oriented Approach to Data Integration} &
  \multicolumn{1}{l|}{Distributed Information Systems in Business} \\ \midrule
\multicolumn{1}{|l|}{6} &
  \multicolumn{1}{l|}{\cite{Preuner98}} &
  \multicolumn{1}{l|}{1998} &
  \multicolumn{1}{l|}{Preuner, G and Schrefl, M} &
  \multicolumn{1}{l|}{Observation consistent integration of views of object life-cycles} &
  \multicolumn{1}{l|}{BNCOD Advances in Databases} \\ \midrule
\multicolumn{1}{|l|}{7} &
  \multicolumn{1}{l|}{\cite{Dijkman06}} &
  \multicolumn{1}{l|}{2006} &
  \multicolumn{1}{l|}{\begin{tabular}[c]{@{}l@{}}Dijkman, R and Ferreira Pires, L\\  and Joosten, S\end{tabular}} &
  \multicolumn{1}{l|}{Integration of heterogeneous BPM Schemas: The Case of XPDL and BPEL} &
  \multicolumn{1}{l|}{The 18th Conference on Advanced Information Systems Engineering} \\ \midrule
\multicolumn{1}{|l|}{8} &
  \multicolumn{1}{l|}{\cite{Stumptner2004}} &
  \multicolumn{1}{l|}{2004} &
  \multicolumn{1}{l|}{\begin{tabular}[c]{@{}l@{}}Stumptner, M and Schrefl, M \\ and Grossmann, G\end{tabular}} &
  \multicolumn{1}{l|}{On the Road to Behavior-Based Integration} &
  \multicolumn{1}{l|}{Proceedings of the first Asian-Pacific conference on Conceptual modelling} \\ \midrule
\multicolumn{1}{|l|}{9} &
  \multicolumn{1}{l|}{\cite{kcpm05}} &
  \multicolumn{1}{l|}{2006} &
  \multicolumn{1}{l|}{Vohringer, J and Mayr, HC} &
  \multicolumn{1}{l|}{Integration of schemas on the pre-design level using the KCPM-approach} &
  \multicolumn{1}{l|}{Advances in Information Systems Development} \\ \midrule
\multicolumn{1}{|l|}{10} &
  \multicolumn{1}{l|}{\cite{Mendling2006}} &
  \multicolumn{1}{l|}{2006} &
  \multicolumn{1}{l|}{Mendling, Jan and Simon, Carlo} &
  \multicolumn{1}{l|}{Business Process Design by View Integration} &
  \multicolumn{1}{l|}{Business Process Management Workshops} \\ \midrule
\multicolumn{1}{|l|}{11} &
  \multicolumn{1}{l|}{\cite{Morrison2009}} &
  \multicolumn{1}{l|}{2009} &
  \multicolumn{1}{l|}{\begin{tabular}[c]{@{}l@{}}Morrison, E and Menzies, A and Koliadis, G\\  and Ghose, A\end{tabular}} &
  \multicolumn{1}{l|}{Business Process Integration: Method and Analysis} &
  \multicolumn{1}{l|}{Proceedings of the Sixth Asia-Pacific Conference on Conceptual Modelling} \\ \midrule
\multicolumn{1}{|l|}{12} &
  \multicolumn{1}{l|}{\cite{Tran2011}} &
  \multicolumn{1}{l|}{2011} &
  \multicolumn{1}{l|}{Tran, H and Zdun, U and Dustdar, S} &
  \multicolumn{1}{l|}{Name-based view integration for enhancing the reusability in process-driven SOAs} &
  \multicolumn{1}{l|}{Business Process Management Workshops} \\ \midrule
\multicolumn{1}{|l|}{13} &
  \multicolumn{1}{l|}{\cite{tran2007}} &
  \multicolumn{1}{l|}{2007} &
  \multicolumn{1}{l|}{Tran, H and Zdun, U and Dustdar, S} &
  \multicolumn{1}{l|}{\begin{tabular}[c]{@{}l@{}}View-based and Model-driven Approach for Reducing the Development Complexity in \\ Process-Driven SOA\end{tabular}} &
  \multicolumn{1}{l|}{International Working Conference on Business Process and Services Computing} \\ \midrule
\multicolumn{1}{|l|}{14} &
  \multicolumn{1}{l|}{\cite{sousa2007}} &
  \multicolumn{1}{l|}{2007} &
  \multicolumn{1}{l|}{\begin{tabular}[c]{@{}l@{}}Sousa, P and Pereira, C and Vendeirinho, R\\  and Caetano, A and Tribolet, J\end{tabular}} &
  \multicolumn{1}{l|}{Applying the Zachman Framework Dimensions to Support Business Process Modeling} &
  \multicolumn{1}{l|}{Digital Enterprise Technology} \\ \midrule
\multicolumn{1}{|l|}{15} &
  \multicolumn{1}{l|}{\cite{colaco2017}} &
  \multicolumn{1}{l|}{2017} &
  \multicolumn{1}{l|}{Colaco, J and Sousa, P} &
  \multicolumn{1}{l|}{View integration of business process models} &
  \multicolumn{1}{l|}{European, Mediterranean, and Middle Eastern Conference on Information Systems} \\ \midrule
\multicolumn{1}{|l|}{16} &
  \multicolumn{1}{l|}{\cite{caetano2012}} &
  \multicolumn{1}{l|}{2012} &
  \multicolumn{1}{l|}{Caetano, A and Pereira, C and Sousa, P} &
  \multicolumn{1}{l|}{Generation of Business Process Model Views} &
  \multicolumn{1}{l|}{\begin{tabular}[c]{@{}l@{}}Conference of ENTERprise Information Systems – aligning technology, \\ organizations and people\end{tabular}} \\ \midrule
\multicolumn{1}{|l|}{17} &
  \multicolumn{1}{l|}{\cite{huang2014}} &
  \multicolumn{1}{l|}{2014} &
  \multicolumn{1}{l|}{Huang, Y and He, K and Feng, Z} &
  \multicolumn{1}{l|}{Business process consolidation based on E-RPSTs} &
  \multicolumn{1}{l|}{IEEE World Congress on Services} \\ \midrule
\multicolumn{1}{|l|}{18} &
  \multicolumn{1}{l|}{\cite{derguech2017}} &
  \multicolumn{1}{l|}{2017} &
  \multicolumn{1}{l|}{Derguech, W and Bhiri, S and Curry, E} &
  \multicolumn{1}{l|}{Designing business capability-aware configurable process models} &
  \multicolumn{1}{l|}{Information Systems} \\ \midrule
\multicolumn{1}{|l|}{19} &
  \multicolumn{1}{l|}{\cite{kunchala2017}} &
  \multicolumn{1}{l|}{2017} &
  \multicolumn{1}{l|}{\begin{tabular}[c]{@{}l@{}}Kunchala, J and Yu, J and Yongchareon, S\\  and Han, Y\end{tabular}} &
  \multicolumn{1}{l|}{Towards merging collaborating processes for artifact lifecycle synthesis} &
  \multicolumn{1}{l|}{Proceedings of the Australasian Computer Science Week Multiconference} \\ \midrule
\multicolumn{1}{|l|}{20} &
  \multicolumn{1}{l|}{\cite{kunchala2019}} &
  \multicolumn{1}{l|}{2019} &
  \multicolumn{1}{l|}{\begin{tabular}[c]{@{}l@{}}Kunchala, J and Yu, J and Yongchareon, S\\  and Liu, C\end{tabular}} &
  \multicolumn{1}{l|}{\begin{tabular}[c]{@{}l@{}}An approach to merge collaborating processes of an inter-organizational business \\ process for artifact lifecycle synthesis\end{tabular}} &
  \multicolumn{1}{l|}{Computing} \\ \midrule
\multicolumn{1}{|l|}{21} &
  \multicolumn{1}{l|}{\cite{rosa2010}} &
  \multicolumn{1}{l|}{2010} &
  \multicolumn{1}{l|}{\begin{tabular}[c]{@{}l@{}}Rosa, L and Dumas, M and Uba, R \\ and Dijkman, R\end{tabular}} &
  \multicolumn{1}{l|}{Merging business process models} &
  \multicolumn{1}{l|}{On the Move to Meaningful Internet Systems} \\ \midrule
\multicolumn{1}{|l|}{22} &
  \multicolumn{1}{l|}{\cite{cardoso2020}} &
  \multicolumn{1}{l|}{2020} &
  \multicolumn{1}{l|}{Cardoso, D and Sousa, P} &
  \multicolumn{1}{l|}{Generation of Stakeholder-Specific BPMN Models} &
  \multicolumn{1}{l|}{Advances in Enterprise Engineering XIII} \\ \midrule
\multicolumn{1}{|l|}{23} &
  \multicolumn{1}{l|}{\cite{Grossmann2005}} &
  \multicolumn{1}{l|}{2005} &
  \multicolumn{1}{l|}{\begin{tabular}[c]{@{}l@{}}Grossmann, G and Ren, Y and Schrefl, M\\  and Stumptner, M\end{tabular}} &
  \multicolumn{1}{l|}{Behavior based integration of composite business processes} &
  \multicolumn{1}{l|}{International Conference on Business Process Management} \\ \midrule
\multicolumn{1}{|l|}{24} &
  \multicolumn{1}{l|}{\cite{Gottschalk2008}} &
  \multicolumn{1}{l|}{2008} &
  \multicolumn{1}{l|}{\begin{tabular}[c]{@{}l@{}}Gottschalk, F and Van Der Aalst, W \\ and Jansen-Vullers, M\end{tabular}} &
  \multicolumn{1}{l|}{Merging event-driven process chains} &
  \multicolumn{1}{l|}{On the Move to Meaningful Internet Systems} \\ \midrule
\multicolumn{1}{|l|}{25} &
  \multicolumn{1}{l|}{\cite{Li2009}} &
  \multicolumn{1}{l|}{2009} &
  \multicolumn{1}{l|}{Li, C and Reichert, M and Wombacher, A} &
  \multicolumn{1}{l|}{Discovering reference models by mining process variants using a heuristic approach} &
  \multicolumn{1}{l|}{Business Process Management} \\ \midrule
\multicolumn{1}{|l|}{26} &
  \multicolumn{1}{l|}{\cite{rosa2013}} &
  \multicolumn{1}{l|}{2013} &
  \multicolumn{1}{l|}{\begin{tabular}[c]{@{}l@{}}La Rosa, M and Dumas, M and Uba, R \\ and Dijkman, R\end{tabular}} &
  \multicolumn{1}{l|}{Business Process Model Merging: An Approach to Business Process Consolidation} &
  \multicolumn{1}{l|}{ACM Transactions on Software Engineering and Methodology} \\ \midrule
\multicolumn{1}{|l|}{27} &
  \multicolumn{1}{l|}{\cite{Assy13}} &
  \multicolumn{1}{l|}{2013} &
  \multicolumn{1}{l|}{Assy, N and Chan, N and Gaaloul, W} &
  \multicolumn{1}{l|}{Assisting business process design with configurable process fragments} &
  \multicolumn{1}{l|}{IEEE International Conference on Services Computing} \\ \midrule
\multicolumn{1}{|l|}{28} &
  \multicolumn{1}{l|}{\cite{kuster2008}} &
  \multicolumn{1}{l|}{2008} &
  \multicolumn{1}{l|}{\begin{tabular}[c]{@{}l@{}}Kuster, J and Gerth, C and Forster, A \\ and Engels, G\end{tabular}} &
  \multicolumn{1}{l|}{Process Merging in Business-Driven Development} &
  \multicolumn{1}{l|}{Proceedings of the Forum at the CAiSE'08} \\ \midrule
\multicolumn{1}{|l|}{29} &
  \multicolumn{1}{l|}{\cite{Nguyen17}} &
  \multicolumn{1}{l|}{2017} &
  \multicolumn{1}{l|}{Nguyen, T and Hong, T and Le Thanh, N} &
  \multicolumn{1}{l|}{\begin{tabular}[c]{@{}l@{}}An ontological approach for organizing a knowledge base to share and reuse business \\ workflow templates\end{tabular}} &
  \multicolumn{1}{l|}{International Conference on Information Science and Technology} \\ \midrule
\multicolumn{1}{|l|}{30} &
  \multicolumn{1}{l|}{\cite{Schumm2010}} &
  \multicolumn{1}{l|}{2010} &
  \multicolumn{1}{l|}{Schumm, D and Leymann, F and Streule, A} &
  \multicolumn{1}{l|}{Process viewing patterns} &
  \multicolumn{1}{l|}{IEEE International Enterprise Distributed Object Computing Conference} \\ \midrule
\multicolumn{1}{|l|}{31} &
  \multicolumn{1}{l|}{\cite{Weidlich2011}} &
  \multicolumn{1}{l|}{2011} &
  \multicolumn{1}{l|}{Weidlich, M and Mendling, J and Weske, M} &
  \multicolumn{1}{l|}{A foundational approach for managing process variability} &
  \multicolumn{1}{l|}{International Conference on Advanced Information Systems Engineering} \\ \midrule
\multicolumn{1}{|l|}{32} &
  \multicolumn{1}{l|}{\cite{kuster2007}} &
  \multicolumn{1}{l|}{2007} &
  \multicolumn{1}{l|}{Kuster, J and Ryndina, K and Gall, H} &
  \multicolumn{1}{l|}{Generation of business process models for object life cycle compliance} &
  \multicolumn{1}{l|}{International Conference on Business Process Management} \\ \midrule
\multicolumn{1}{|l|}{33} &
  \multicolumn{1}{l|}{\cite{zachaman2010}} &
  \multicolumn{1}{l|}{2010} &
  \multicolumn{1}{l|}{Sowa, J and Zachman, J} &
  \multicolumn{1}{l|}{Extending and formalizing the framework for information systems architecture} &
  \multicolumn{1}{l|}{IBM Systems Journal} \\ \midrule
\multicolumn{1}{|l|}{34} &
  \multicolumn{1}{l|}{\cite{pereira2011}} &
  \multicolumn{1}{l|}{2011} &
  \multicolumn{1}{l|}{Pereira, C and Caetano, A and Sousa, P} &
  \multicolumn{1}{l|}{Ontology-Driven Business Process Design} &
  \multicolumn{1}{l|}{Conference on e-Business, e-Services and e-Society} \\ \midrule
\multicolumn{1}{|l|}{35} &
  \multicolumn{1}{l|}{\cite{Weidlich12}} &
  \multicolumn{1}{l|}{2012} &
  \multicolumn{1}{l|}{Weidlich, M and Mendling, J} &
  \multicolumn{1}{l|}{Perceived consistency between process models} &
  \multicolumn{1}{l|}{Information Systems} \\ \midrule
\multicolumn{1}{|l|}{36} &
  \multicolumn{1}{l|}{\cite{schewe2015}} &
  \multicolumn{1}{l|}{2015} &
  \multicolumn{1}{l|}{Schewe et al.} &
  \multicolumn{1}{l|}{Horizontal business process model integration} &
  \multicolumn{1}{l|}{Transactions on Large-Scale Data- and Knowledge-Centered Systems XVIII} \\ \midrule
\multicolumn{1}{|l|}{37} &
  \multicolumn{1}{l|}{\cite{milani2015}} &
  \multicolumn{1}{l|}{2015} &
  \multicolumn{1}{l|}{\begin{tabular}[c]{@{}l@{}}Milani, F and Dumas, M and Matulevivcius, R\\  and Ahmed, N and Kasela, S\end{tabular}} &
  \multicolumn{1}{l|}{\begin{tabular}[c]{@{}l@{}}Criteria and heuristics for business process model decomposition: Review \\ and comparative evaluation\end{tabular}} &
  \multicolumn{1}{l|}{Business and Information Systems Engineering} \\ \midrule
\multicolumn{1}{|l|}{38} &
  \multicolumn{1}{l|}{\cite{pini2015}} &
  \multicolumn{1}{l|}{2015} &
  \multicolumn{1}{l|}{Pini, A and Brown, R} &
  \multicolumn{1}{l|}{Process visualization techniques for multi-perspective process comparisons} &
  \multicolumn{1}{l|}{Asia-Pacific Conference on Business Process Management} \\ \midrule
\multicolumn{1}{|l|}{39} &
  \multicolumn{1}{l|}{\cite{apromore2009}} &
  \multicolumn{1}{l|}{2009} &
  \multicolumn{1}{l|}{Rosa et al.} &
  \multicolumn{1}{l|}{APROMORE: An advanced process model repository} &
  \multicolumn{1}{l|}{Expert Systems with Applications} \\ \midrule
\multicolumn{1}{|l|}{40} &
  \multicolumn{1}{l|}{\cite{Belchior2020}} &
  \multicolumn{1}{l|}{2020} &
  \multicolumn{1}{l|}{\begin{tabular}[c]{@{}l@{}}Belchior, R and Vasconcelos, A \\ and Guerreiro, S and Correia, M\end{tabular}} &
  \multicolumn{1}{l|}{A Survey on Blockchain Interoperability: Past} &
  \multicolumn{1}{l|}{arXiv} \\ \midrule
\multicolumn{1}{|l|}{41} &
  \multicolumn{1}{l|}{\cite{bitcoin}} &
  \multicolumn{1}{l|}{2008} &
  \multicolumn{1}{l|}{Nakamoto, S} &
  \multicolumn{1}{l|}{Bitcoin: A peer-to-peer electronic cash system} &
  \multicolumn{1}{l|}{Available online} \\ \midrule
\multicolumn{1}{|l|}{42} &
  \multicolumn{1}{l|}{\cite{fabric}} &
  \multicolumn{1}{l|}{2018} &
  \multicolumn{1}{l|}{Androulaki et al.} &
  \multicolumn{1}{l|}{Hyperledger Fabric: A Distributed Operating System for Permissioned Blockchains} &
  \multicolumn{1}{l|}{EuroSys} \\ \midrule
\multicolumn{1}{|l|}{43} &
  \multicolumn{1}{l|}{\cite{eu19}} &
  \multicolumn{1}{l|}{2019} &
  \multicolumn{1}{l|}{\begin{tabular}[c]{@{}l@{}}European Parliament and \\ European Council\end{tabular}} &
  \multicolumn{1}{l|}{\begin{tabular}[c]{@{}l@{}}Blockchain and the General Data Protection Regulation Can distributed ledgers be \\ squared with European data protection law?\end{tabular}} &
  \multicolumn{1}{l|}{Available online} \\ \midrule
\multicolumn{1}{|l|}{44} &
  \multicolumn{1}{l|}{\cite{audits2020}} &
  \multicolumn{1}{l|}{2018} &
  \multicolumn{1}{l|}{KPMG} &
  \multicolumn{1}{l|}{Auditing Blockchain Solutions} &
  \multicolumn{1}{l|}{Available online} \\ \midrule
\multicolumn{1}{|l|}{45} &
  \multicolumn{1}{l|}{\cite{pereira2011b}} &
  \multicolumn{1}{l|}{2011} &
  \multicolumn{1}{l|}{\begin{tabular}[c]{@{}l@{}}Marques Pereira, C and Caetano, A\\  and Sousa, P\end{tabular}} &
  \multicolumn{1}{l|}{Using a controlled vocabulary to support business process design} &
  \multicolumn{1}{l|}{Workshop on Enterprise and Organizational Modeling and Simulation} \\ \midrule
\multicolumn{1}{|l|}{46} &
  \multicolumn{1}{l|}{\cite{navathe82}} &
  \multicolumn{1}{l|}{1982} &
  \multicolumn{1}{l|}{Navathe, S and Gadgil, S} &
  \multicolumn{1}{l|}{A Methodology for View Inegration in Logical Database Design} &
  \multicolumn{1}{l|}{International Conference on Very Large Databases} \\ \midrule
\multicolumn{1}{|l|}{47} &
  \multicolumn{1}{l|}{\cite{pereira2011}} &
  \multicolumn{1}{l|}{2011} &
  \multicolumn{1}{l|}{Pereira, C and Caetano, A and Sousa, P} &
  \multicolumn{1}{l|}{Ontology-Driven Business Process Design} &
  \multicolumn{1}{l|}{Conference on e-Business, e-Services and e-Society} \\ \midrule
\multicolumn{1}{|l|}{48} &
  \multicolumn{1}{l|}{\cite{sousa2019}} &
  \multicolumn{1}{l|}{2019} &
  \multicolumn{1}{l|}{Sousa, P and Cardoso, D and Colaco, J} &
  \multicolumn{1}{l|}{Managing Multi-view Business Processes Models in the Atlas Tool} &
  \multicolumn{1}{l|}{Forum/Posters/CIAO!DC@EEWC 2019} \\ \midrule
\multicolumn{1}{|l|}{49} &
  \multicolumn{1}{l|}{\cite{rosa2010}} &
  \multicolumn{1}{l|}{2010} &
  \multicolumn{1}{l|}{\begin{tabular}[c]{@{}l@{}}Rosa, L and Dumas, M and Uba, R \\ and Dijkman, R\end{tabular}} &
  \multicolumn{1}{l|}{Merging business process models} &
  \multicolumn{1}{l|}{On the Move to Meaningful Internet Systems} \\ \midrule
\multicolumn{1}{|l|}{50} &
  \multicolumn{1}{l|}{\cite{hyperldegerpd}} &
  \multicolumn{1}{l|}{2020} &
  \multicolumn{1}{l|}{Hyperledger Foundation} &
  \multicolumn{1}{l|}{Hyperledger Fabric Private Data} &
  \multicolumn{1}{l|}{Available online} \\ \midrule
\multicolumn{1}{|l|}{51} &
  \multicolumn{1}{l|}{\cite{belchior2019_audits}} &
  \multicolumn{1}{l|}{2020} &
  \multicolumn{1}{l|}{\begin{tabular}[c]{@{}l@{}}Belchior, R and Vasconcelos, A \\ and Correia, M\end{tabular}} &
  \multicolumn{1}{l|}{Towards Secure Decentralized  and Automatic Audits with Blockchain} &
  \multicolumn{1}{l|}{European Conference on Information Systems} \\ \bottomrule
\end{tabular}
}
\end{table}

\end{document}